\documentclass[pra,showpacs,floatfix]{revtex4}
\usepackage{amsmath}
\usepackage{amssymb}
\usepackage{graphicx}

\begin{document}

\title{Discrete localized modes supported by an inhomogeneous defocusing
nonlinearity}
\author{Goran Gligori\'{c}$^{1}$, Aleksandra Maluckov$^{1}$, Ljup\v co Had%
\v{z}ievski$^{1}$, and Boris A. Malomed$^{2}$} \affiliation{$^1$
P$^{}$ group, Vin\v ca Institute of Nuclear Sciences, University of
Belgrade, P. O. B. 522,11001 Belgrade, Serbia\\
$^{2}$ Department of Physical Electronics, School of Electrical Engineering,
Faculty of Engineering, Tel Aviv University, Tel Aviv 69978, Israel}

\begin{abstract}
We report that infinite and semi-infinite lattices with spatially
inhomogeneous self-defocusing (SDF)\ onsite nonlinearity, whose strength
increases rapidly enough toward the lattice periphery, support stable
unstaggered (UnST) discrete bright solitons, which do not exist in lattices
with the spatially uniform SDF nonlinearity. The UnST solitons coexist with
stable staggered (ST) localized modes, which are always possible under the
defocusing onsite nonlinearity. The results are obtained in a numerical
form, and also by means of variational approximation (VA). In the
semi-infinite (truncated) system, some solutions for the UnST surface
solitons are produced in an exact form. On the contrary to surface discrete
solitons in uniform truncated lattices, the threshold value of the norm
vanishes for the UnST solitons in the present system. Stability regions for
the novel UnST solitons are identified. The same results imply the existence
of ST discrete solitons in lattices with the spatially growing self-focusing
nonlinearity, where such solitons cannot exist either if the nonlinearity is
homogeneous. In addition, a lattice with the uniform onsite SDF nonlinearity
and exponentially decaying inter-site coupling is introduced and briefly
considered too. Via a similar mechanism, it may also support UnST discrete
solitons, under the action of the SDF nonlinearity. The results may be
realized in arrayed optical waveguides and collisionally inhomogeneous
Bose-Einstein condensates trapped in deep optical lattices. A generalization
for a two-dimensional system is briefly considered too.
\end{abstract}

\maketitle

\section{Introduction}

The studies of self-sustained localized modes (bright solitons) in various
physical settings have demonstrated that they can be supported by the
self-focusing nonlinearity \cite{1}, or, in the form of gap solitons, by the
self-defocusing (SDF)\ nonlinearity combined with periodic linear potentials
\cite{2}, or in systems where the nonlinearity periodically changes its
magnitude, and even the sign, along the evolution variable or in the
transverse directions \cite{3}, \cite{Barcelona_review}. Guiding bright
solitons by pure SDF nonlinearities, without the help of a linear potential,
was considered obviously impossible, until it was recently demonstrated in
Refs. \cite{Barcelona_slow}- \cite{Zeng} that this is possible if the
strength of the local SDF nonlinearity grows fast enough in space from the
center to periphery, as a function of the radial coordinate, $r$. The
existence of bright solitons in this case is a consequence of the fact that
the growth of the nonlinearity coefficient makes the underlying equations
\emph{nonlinearizable} for the decaying tails (hence, the superposition
principle is not valid for them), i.e., it \ makes unnecessary placing the
propagation constant of the soliton into the semi-infinite spectral gap of
the linearized system, where SDF nonlinearities cannot support any
self-localization. A similar spatially inhomogeneous setting may support
bright solitons under the action of the nonlocal SDF nonlinearity \cite{He}.
Independently, in Ref. \cite{Zhong} it was demonstrated that the uniform SDF
nonlinearity may support bright solitons in a more exotic model, where the
local diffraction coefficient decays at $r\rightarrow \infty $. Although the
latter model is not equivalent to those introduced in Refs. \cite%
{Barcelona_slow} and \cite{Barcelona}, the mechanisms supporting
bright solitons in these ``nonorthodox" settings are similar.

The main subject of the present work is a possibility to create bright
discrete solitons in two distinct one-dimensional (1D) lattice settings with
inhomogeneous SDF onsite nonlinearities. First is a discrete counterpart of
the system introduced in Refs. \cite{Barcelona,Barcelona_slow}, which is
based on the discrete nonlinear-Schr\"{o}dinger (DNLS) /Gross-Pitaevskii
equation for the field amplitude in optical media, or the mean-field wave
function of a Bose-Einstein condensate (BEC). As reviewed in Ref. \cite%
{Barcelona_review}, spatially inhomogeneous nonlinearities can be realized
in various ways in optics \cite{5}. For example, in photorefractive
materials (such as LiNbO$_{3}$) nonuniform doping with Cu or Fe, which
provide for the two-photon resonance, may be used for this purpose \cite{6}.
In the BEC, spatially modulated nonlinearity can be created, via the
Feshbach resonance, by nonuniform external fields \cite{7,8}. Our goal is to
construct discrete localized modes, of both staggered (ST) and unstaggered
(UnST) types, in this setting, and investigate their stability and dynamics.
Only the former type exists and may be stable in discrete media with the SDF
uniform onsite nonlinearity. Therefore, we chiefly focus on stable UnST
localized modes, the existence of which is the most nontrivial finding. In
fact, the staggering transformation makes these modes tantamount to ST ones
in the lattice with the inhomogeneous focusing nonlinearity, which is a
counter-intuitive result too, as such discrete solitons do not exist in
uniform lattices \cite{Kevrekidis}.

The second setting is the semi-infinite lattice with the SDF inhomogeneous
nonlinearity. This discrete system generically supports surface solitons of
the ST type, similarly to the periodic truncated potentials in solid state
media, where staggered localized modes pinned to the surface were first
realized as localized electronic modes (the Tamm states) in solid-state
media \cite{12}, \cite{13}. The self-trapping of light near the edge of a
waveguide array with self-focusing nonlinearity can lead to the formation of
discrete UnST surface solitons \cite{molina}, \cite{10}, \cite{14}. It has
been found that the surface modes acquire novel properties, different from
those of discrete solitons in infinite lattices, such as a threshold power
(norm) necessary for the existence of the surface solitons, and a
possibility of coexistence of two surface modes, stable and unstable ones,
at the same power. Generally, the existence and stability of diverse
localized surface modes in semi-infinite nonlinear lattices are the result
of the interplay between the nonlinearity and discreteness of the array, on
the one hand, and the presence of the edge in the lattice, on the other. In
this context, our goal is to investigate a possibility to generate different
types of surface bright discrete solitons, and analyze their stability and
dynamical properties, in the case when the local SDF strength grows from the
edge into the depth of the lattice. In this case too, we are chiefly
interested in the surface modes of the UnST type, which are impossible at
the edge of a uniform truncated lattice with the SDF onsite nonlinearity.

We also introduce and briefly consider a discrete counterpart of the
above-mentioned model with the uniform SDF nonlinearity and decaying
diffraction coefficient \cite{Zhong}, which corresponds to the inter-site
coupling constant in the lattice, exponentially decaying with the increase
of the discrete coordinate, $|n|$. Unlike the continuous medium, where it is
difficult to realize a decreasing diffraction coefficient, in the lattice it
merely implies a gradually growing spacing between the sites, as the
coupling constant depends on it exponentially. While this model is very
different from the one with the growing onsite SDF nonlinearity, UnST
solitons, which are impossible in the uniform setting, are supported in it
by a similar mechanism.

The paper is organized as follows. In Section II the models are introduced.
The existence and stability of the UnST and ST discrete solitons in the
infinite inhomogeneous lattice are reported in Section III. In addition to
numerical results, this section also presents a variational approximation
(VA) developed in an analytical form. Numerical and analytical results,
including particular exact solutions and the VA, for the UnST and ST surface
solitons in the truncated lattice are reported in Section IV. The paper is
concluded by Section V, where, in particular, we discuss a two-dimensional
generalization of the system, and give some preliminary results for it too.

\section{The models}

\subsection{The infinite lattice}

We here introduce two 1D discrete models in the form of the DNLS
equations in the infinite and semi-infinite lattices (waveguide
arrays, in terms of optics). The SDF onsite nonlinearity is assumed
to grow exponentially from the central site in both directions in
the infinite lattice, and from the edge towards the bulk in the
semi-infinite one. It is necessary to stress that the steep growth
of the local nonlinearity does not imply that one should use, for
instance, a steeply growing density of the dopants (in the optical
realization of the system). Instead, it is enough to assume that the
density is uniform across the lattice, but the detuning of the
two-photon resonance is gradually reduced from large to small values
by means of a slightly inhomogeneous mechanical stress, or by means
of the Zeeman or Stark shift. In any case, this mode of the
resonance adjustment strongly affects only the local nonlinearity,
but does not introduce a conspicuous linear potential, hence effects
of the nonlinearity modulation may be studied in the ``pure" form
\cite{Barcelona}.

It is commonly known that lattices with the homogeneous SDF nonlinearity,
both infinite and semi-infinite ones, give rise to bright discrete solitons
of the ST type, while UnST localized modes cannot be created in such
lattices \cite{Kevrekidis}. In comparison with the infinite lattice, the
truncation introduces an effective repulsive potential for discrete
solitons, which acts in the combination with the periodic (Peierls--Nabarro
\cite{cembel}, \cite{wepn}) potential induced by the lattice. As a result,
ST surface solitons are created if their norm (power) exceeds a certain
threshold value, at which the discreteness may overcome the repulsion from
the surface. As we demonstrate below, properties of the ST discrete
solitons, in the infinite and truncated lattices alike, are not affected
dramatically by the spatial modulation of the nonlinearity, while it opens
the way to the creation of a completely novel species of UnST discrete
solitons.

The 1D discrete model that we consider here is based on the following DNLS
equation with the exponentially growing onsite SDF nonlinearity for complex
field amplitudes $u_{n}$:
\begin{equation}
i\frac{du_{n}}{dz}=-\frac{1}{2}\left( u_{n+1}+u_{n-1}-2u_{n}\right)
+e^{\alpha |n|}\left\vert u_{n}\right\vert ^{2}u_{n},  \label{eq1}
\end{equation}%
where $\alpha \geq 0$ is the growth rate. Stationary solutions to Eqs. (\ref%
{eq1}) with real propagation constant $K$ are looked for as
\begin{equation}
u_{n}(z)=e^{iKz}U_{n},  \label{uU}
\end{equation}%
where real discrete function $U_{n}$ obeys the following equation:
\begin{equation}
-\left( K+1\right) U_{n}=-\frac{1}{2}\left( U_{n+1}+U_{n-1}\right)
+e^{\alpha |n|}U_{n}^{3},  \label{eq2}
\end{equation}%
which can be derived from the Lagrangian,%
\begin{equation}
L=\frac{1}{2}\sum_{n=-\infty }^{+\infty }\left[ \left( K+1\right)
U_{n}^{2}-U_{n}U_{n+1}+\frac{1}{2}e^{\alpha |n|}U_{n}^{4}\right] .  \label{L}
\end{equation}

Note that Eq. (\ref{eq2}) has an exact analytical solution, which is valid
at
\begin{equation}
K<K_{\mathrm{cutoff}}\equiv \cosh \left( \alpha /2\right) -1,  \label{k0}
\end{equation}%
for either $n>0$ or $n<0$ (but not at positive and negative values of $n$
simultaneously, therefore it does not correspond to solitons):%
\begin{eqnarray}
\left( U_{n}\right) _{\mathrm{exact}} &=&A_{\mathrm{exact}}e^{-\alpha |n|/2},
\label{exact} \\
A_{\mathrm{exact}}^{2} &=&\cosh \left( \alpha /2\right) -1-K.  \label{Aexact}
\end{eqnarray}%
In fact, this solution represents the above-mentioned \textit{nonlinearizable%
} \textit{tail} of the discrete solitons. On the other hand, for large $%
\alpha $ and/or large $-K$, the UnST soliton is strongly squeezed around $n=0
$, and can be described by the following truncated approximation:%
\begin{equation}
U_{0}\approx \sqrt{-\left( K+1\right) },U_{\pm 1}\approx \left( -\frac{K+1}{4%
}\right) ^{1/6}e^{-\alpha /3},U_{\pm 2}\approx \left( -\frac{K+1}{2^{8}}%
\right) ^{1/18}e^{-7\alpha /9},  \label{0+1-1}
\end{equation}%
which is valid (for $K+1<0$) under the condition of
\begin{equation}
-\left( K+1\right) e^{\alpha }\gg 1.  \label{>>}
\end{equation}

It is worthy to note that, in the continuum counterpart of Eq. (\ref{eq2}),
with $U_{n+1}+U_{n-1}-2U_{n}$ replaced by $d^{2}U/dx^{2}$ ($x$ is the
continuous coordinate), taking the equation at the inflexion point, where $%
d^{2}U/dx^{2}=0$ (obviously, any continuum-soliton profile features such a
point), proves that the solitons may only exist with $K<0$ \cite{Barcelona}.
Although this proof does not apply to the discrete model, the actual result
is that the discrete UnST solitons too exist only at $K<0$, see Eq. (\ref%
{cutoff}) and Fig. \ref{poredjenje}(b) below.

\subsection{The semi-infinite lattice}

In the truncated (semi-infinite) version of the model, Eq. (\ref{eq1}) is
modified as follows:
\begin{eqnarray}
i\frac{du_{n}}{dz} &=&-\frac{1}{2}\left( u_{n+1}+u_{n-1}-2u_{n}\right)
+e^{\alpha |n|}\left\vert u_{n}\right\vert ^{2}u_{n},~\mathrm{at}~~n\geq 2,
\label{m} \\
i\frac{du_{1}}{dz} &=&-\frac{1}{2}\left( u_{2}+C_{0}u_{0}-2u_{1}\right)
+e^{\alpha }\left\vert u_{1}\right\vert ^{2}u_{1},~\mathrm{at}~~n=1,
\label{1} \\
i\frac{du_{0}}{dz} &=&-\frac{1}{2}C_{0}u_{1}+u_{0}+\sigma _{0}\left\vert
u_{0}\right\vert ^{2}u_{0},~\mathrm{at}~~n=0.  \label{0}
\end{eqnarray}%
It is assumed here that the edge site of the semi-infinite lattice is
labeled by $n=0$, while the nonlinearity at this site may be different from
that in the bulk ($\sigma _{0}\neq 1$), and the coefficient of the
inter-site coupling between the edge and the rest of the lattice may be
different too from its bulk counterpart (which is scaled to be $1$), i.e., $%
C_{0}\neq 1$ in the general case.

The substitution of expression (\ref{uU}) transforms Eqs. (\ref{m})-(\ref{0}%
) into the stationary form,%
\begin{eqnarray}
-\left( K+1\right) U_{n} &=&-\frac{1}{2}\left( U_{n+1}+U_{n-1}\right)
+e^{\alpha |n|}U_{n}^{3},~\mathrm{at}~~n\geq 2,  \label{m1} \\
-\left( K+1\right) U_{1} &=&-\frac{1}{2}\left( U_{2}+C_{0}U_{0}\right)
+e^{\alpha }U_{1}^{3},~\mathrm{at}~~n=1,  \label{11} \\
-\left( K+1\right) U_{0} &=&-\frac{1}{2}C_{0}U_{1}+\sigma _{0}U_{0}^{3},~%
\mathrm{at}~~n=0.  \label{01}
\end{eqnarray}%
Under condition (\ref{>>}), these equations give rise to the same
approximate (strongly squeezed) solution as given by Eq. (\ref{0+1-1}), with
a difference that components $U_{-1}$ and $U_{-2}$ are absent.

Equations (\ref{m1})-(\ref{01}) can be derived from the respective
Lagrangian, cf. Eq. (\ref{L}),%
\begin{gather}
L=\frac{1}{2}\left\{ \sum_{n=1}^{\infty }\left[ \left( K+1\right)
U_{n}^{2}-U_{n}U_{n+1}+\frac{1}{2}e^{\alpha n}U_{n}^{4}\right] \right.
\notag \\
\left. +\left( K+1\right) U_{0}^{2}-C_{0}U_{0}U_{1}+\frac{\sigma _{0}}{2}%
U_{0}^{4}\right\} ,  \label{lagrangian}
\end{gather}%
The total power (norm) of the modes in the infinite and truncated lattices
is defined, respectively, as%
\begin{equation}
P_{\mathrm{infin}}=\sum_{n=-\infty }^{+\infty }U_{n}^{2},~P_{\mathrm{trunc}%
}=\sum_{n=0}^{+\infty }U_{n}^{2}.  \label{P}
\end{equation}%
It is relevant to mention that obvious rescaling,
\begin{equation}
\left( U_{n}\right) _{\mathrm{infin}}\equiv \mathrm{sign}(n)\cdot e^{\alpha
/2}\left( U_{|n|-1}\right) _{\mathrm{trunc}},~\mathrm{for~}~|n|~\geq 1,
\label{inf-tru}
\end{equation}%
and $\left( U_{0}\right) _{\mathrm{infin}}=0$, of discrete solitons found in
the truncated lattice with $C_{0}=\sigma _{0}=1$ yields odd (alias "twisted"
\cite{twist,Kevrekidis}) soliton modes in the infinite lattice.

As said above, our objective is to use the inhomogeneous SDF nonlinearity
for creating stable discrete localized modes, both in the bulk and at the
surface, as suggested by the analysis recently reported for continuous
models \cite{Barcelona}. In most cases, such modes, both fundamental and
topological ones (multipoles and vortices) are stable, and unstable ones
transform into tightly confined breathers.

The bright solitons in the continuous models were found also with the growth
rate of the nonlinearity slower than exponential. The necessary and
sufficient condition for supporting solitons with a finite norm by the SDF
nonlinearity in the $D$-dimensional continuous space is that the
nonlinearity coefficient in front of the cubic term, $g(r)$ [such as $%
e^{\alpha |n|}$ in Eq. (\ref{eq1})], must grow with the distance from the
center, $r$, faster than $r^{D}$:%
\begin{equation}
\lim_{r\rightarrow \infty }\left[ r^{D}/g(r)\right] =0.  \label{D}
\end{equation}%
The same condition remains true in discrete systems with the growing SDF
nonlinearity, because the condition is determined by the asymptotic form of
the solution far from the center, where the discrete medium seems as a
quasi-continuum. Furthermore, condition (\ref{D}) suggests not only the
existence of the normalizable self-trapped modes, but also the fact that
they may realize the ground state of the system, for given norm $P$. Indeed,
the energy (Hamiltonian $H$) of such modes is obviously positive, while a
competing trivial state, which might presumably give rise to $H=0$, thus
demonstrating that the self-trapped modes do not represent the ground state,
may be taken as a delocalized distribution with radius $R\rightarrow \infty $
and vanishing density
\begin{equation}
U_{\mathrm{deloc}}^{2}=\alpha _{D}^{-1}R^{-D}P,  \label{deloc}
\end{equation}%
where the volume coefficient is $\alpha _{1}=2,\alpha _{2}=\pi $, and $%
\alpha _{3}=(4/3)\pi $. The corresponding quartic term in the Hamiltonian
density is $h_{\mathrm{quart}}=(1/2)g(r)U_{\mathrm{deloc}}^{4}$, giving rise
to a term in $H$ which is estimated as
\begin{equation}
H_{\mathrm{quart}}^{\mathrm{(deloc)}}=\frac{\alpha _{D}}{4}U_{\mathrm{deloc}%
}^{4}\int_{0}^{R}g(r)r^{D-1}dr.  \label{H}
\end{equation}%
Finally, the substitution of Eq. (\ref{deloc}) into this expression
demonstrates that $H_{\mathrm{quart}}^{\mathrm{(deloc)}}$\emph{\ diverges}
in the limit of $R\rightarrow \infty $ exactly in the case when condition (%
\ref{D}) holds, hence the delocalized state cannot compete with the
self-trapped one in the selection of the ground state.

The advantage of using the exponential \ modulation is a possibility to find
particular solutions in an exact analytical form, and accurate results
produced in this case by the VA \cite{Barcelona_slow,Zeng}. Therefore, here
too we consider the model with the exponentially modulated nonlinearity, as
already fixed in Eqs. (\ref{eq1}) and\ (\ref{m}). Both particular exact
solutions and VA-based predictions are produced below. On the other hand, it
is relevant to stress that the results are definitely structurally stable
against a change of the particular form of the nonlinearity modulation, the
only condition being that its local strength must not grow slower than
prescribed by Eq. (\ref{D}).

\subsection{The lattice with the exponentially decaying coupling constant}

Let us briefly mention the other possibility to obtain UnST
solitons in the lattices with SDF nonlinearity. These lattices are
characterized by the uniform SDF onsite nonlinearity and
exponentially decreasing constant of the inter-site coupling. The
model equations can be derived from the following
Lagrangian of the inter-site coupling, cf. Eqs. (\ref{L}) and (\ref%
{lagrangian}):
\begin{equation}
L_{\mathrm{coupl}}^{\mathrm{(\exp -decay)}}=\frac{1}{2}\sum_{n}\left(
e^{-\alpha \left\vert n-1\right\vert }u_{n}^{\ast }u_{n-1}+e^{-\alpha
\left\vert n\right\vert }u_{n}^{\ast }u_{n+1}\right) .  \label{Lcouple}
\end{equation}%
The respective DNLS equation for stationary solutions, sought for
in the form of Eq. (\ref{uU}), is
\begin{equation}
-KU_{n}=-\frac{1}{2}\left( e^{-\alpha \left\vert n-1\right\vert
}U_{n-1}+e^{-\alpha \left\vert n\right\vert }U_{n+1}\right) +U_{n}^{3}~.
\label{K}
\end{equation}%
In particular, Eq. (\ref{K}) admits an exact UnST exponential solution,
which, as well as its counterpart (\ref{exact}) considered above, does not
represent a soliton, and is a less generic solution, as it can be found in
the exact form solely for $K=0$:%
\begin{gather}
K=0,~U_{n}=\tilde{A}_{\mathrm{exact}}e^{-\alpha |n|/2},  \label{tilde} \\
\left( \tilde{A}_{\mathrm{exact}}\right) ^{2}=e^{\alpha /2}\cosh \left(
\alpha \right) .  \label{Atilde}
\end{gather}

This result can be used to construct an exact solution for a
discrete UnST soliton pinned to the edge to the truncated version of
the present lattice, occupying the region of $n\geq 0$, with an
``anomaly", accounted for by coefficient $C_{0}\neq 1$, in the
constant of the coupling
between the edge site, $n=0$, and the neighboring one, $n=1$, cf. Eqs. (\ref%
{1}) and (\ref{0}) [unlike Eq. (\ref{0}), we here assume that the onsite
nonlinearity is completely uniform, i.e., $\sigma _{0}\equiv 1$]. A simple
analysis shows that the exact solution for the pinned soliton is given by
Eqs. (\ref{tilde}) and (\ref{Atilde}) at $n>0$, while%
\begin{equation}
U_{0}=\left( C_{0}\tilde{A}_{\mathrm{exact}}/2\right) ^{1/3}  \label{U0tilde}
\end{equation}%
at $n=0$, provided that $C_{0}$ takes the special value,%
\begin{equation}
C_{0}^{4}=2e^{\alpha /2}\cosh \left( \alpha \right) .  \label{C0}
\end{equation}%
Note that this value exceeds its bulk counterpart, $C_{0}>\left(
C_{0}\right) _{\mathrm{bulk}}\equiv 1$.

For the truncated lattice with the exponentially growing onsite
nonlinearity, which is described by Eqs. (\ref{m})-(\ref{0}), a similar
exact solution is given below, see Eqs. (\ref{Um})-(\ref{A^2}). Although
such exact solutions are not generic ones, they provide the rigorous proof
of the existence of the UnST discrete solitons in the lattices with the SDF
nonlinearity, where solitons of this type cannot normally exist.

\section{The infinite lattice}

\subsection{Numerical results}

For the numerical solution of the stationary equations, Eqs. (\ref{eq2}) and
(\ref{m1})-(\ref{01}), we adopted the nonlinear equation solver based on the
Powell method \cite{nashi}. Direct dynamical simulations of dynamical
equations (\ref{eq1}) and (\ref{m})-(\ref{0}) were based on the Runge-Kutta
numerical procedure of the sixth order.

\begin{figure}[h]
\center\includegraphics [width=13cm]{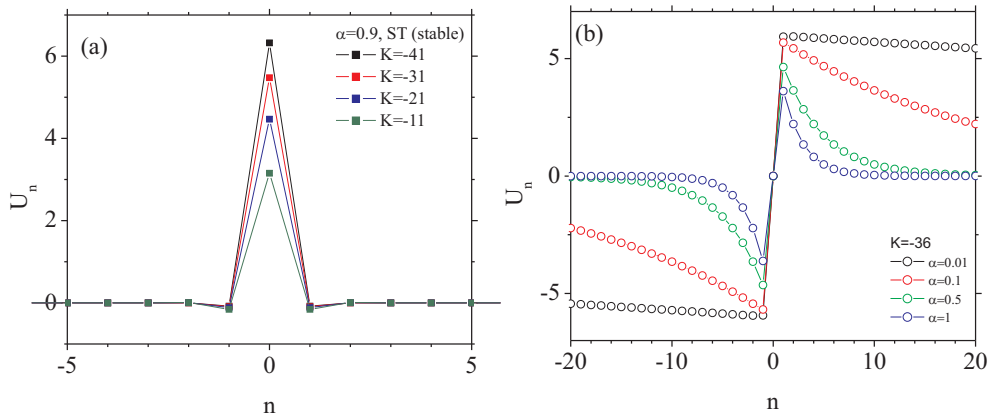}
\caption{(Color online) Profiles of stable staggered (a) and unstable
twisted solitons (b), produced by the numerical calculations for parameters
indicated in the figure.}
\label{f1}
\end{figure}

It is well known that the lattice with the uniform onsite SDF nonlinearity
supports stable bright solitons of the ST type, in onsite and inter-site
configurations \cite{Kevrekidis}. The ST soliton branches survive the change
of the nonlinearity from uniform to exponentially modified, as in Eq. (\ref%
{eq1}). Numerical calculations have shown that accordingly modified ST
solitons, strongly pinned to the center of the lattice, can be found at all
values of $\alpha >0$, see Fig. \ref{f1}(a). These discrete solitons are
always stable.

Our main objective here is to demonstrate that the inhomogeneous onsite SDF
nonlinearity supports stable UnST localized modes. In lattices with the
uniform self-focusing nonlinearity, such solutions can be realized in two
configurations with respect to the position of the soliton's center, onsite
and inter-site, i.e., centered on lattice site, or between two adjacent
lattice sites, respectively. In the present model, the center of the UnST
solitons naturally coincides with the minimum of the onsite nonlinearity
strength (the bottom point of the U profile).

\begin{figure}[h]
\center\includegraphics [width=13cm]{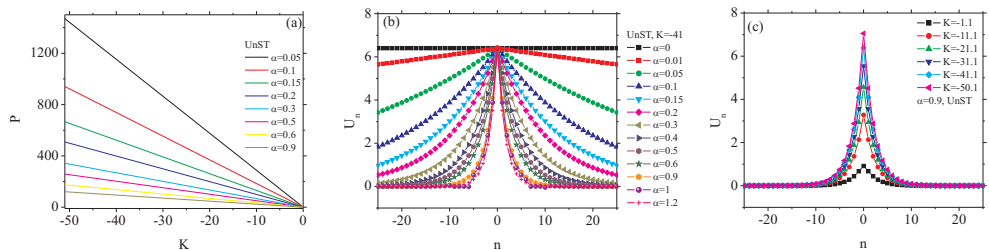}
\caption{(Color online) (a) The power versus the propagation constant for
unstaggered solitons. Plots (b) and (c) display shapes of the solitons. The
respective values of $\protect\alpha $ and $K$ are indicated in the panels.}
\label{f2}
\end{figure}

In Figs. \ref{f2}(a) and \ref{f2}(b), dependencies of the total power on the
propagation constant, and amplitude profiles are shown for UnST solitons at
different values of steepness $\alpha $ of the nonlinearity-modulation
profile in Eq. (\ref{eq1}). Shapes of the modes for a fixed $\alpha $ and
different values of propagation constant $K$ are displayed in Fig. \ref{f2}%
(c). In the lattice with the uniform onsite SDF nonlinearity, UnST localized
modes do not exist (and, respectively, the uniform UnST background is not
subject to the modulational instability). The exponential growth of the
onsite nonlinearity strength from the center to periphery gives rise to UnST
modes. At very small values of the growth rate, $\alpha $, the mode seems as
the background with a weak maximum at the lattice center, $n=0$. The pinned
UnST localized modes, with negligible tail amplitudes at the periphery of
the modulated lattice, are found at $\alpha >$ $\alpha _{\min }\approx 0.1$,
when the exponential modulation is steep enough to trap the UnST soliton at
the center, in the framework of the present numerical scheme.

\begin{figure}[h]
\center\includegraphics [width=13cm]{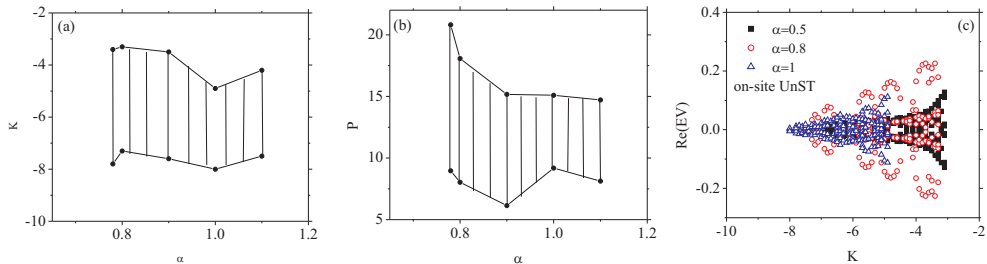}
\caption{The unstaggered solitons are subject to an oscillatory instability
in the hatched areas, which are shown in the planes of $\left( K,\protect%
\alpha \right) $ and $\left( P,\protect\alpha \right) $ in plots (a)
and (b), respectively. Examples of sets of real parts of complex
eigenvalues (``EVs") of perturbation modes are plotted versus $K$ in
(c). The instability is accounted for by $\mathrm{Re(EV)}>0.$
Parameter values are indicated in the panels.} \label{f3}
\end{figure}

The UnST soliton solutions have been found numerically at $\alpha <\alpha
_{\max }\approx 1.5$. This limit is imposed by the numerical scheme, but not
by the system per se [indeed, Eq. (\ref{0+1-1}) demonstrates the existence
of solutions at large values of $\alpha$]. The linear-stability analysis
predicts that the newly found UnST discrete solitons are chiefly stable in
the their existence region, except for a relatively narrow area shown in
Fig. \ref{f3}, where they are subject to an oscillatory instability.

\begin{figure}[h]
\center\includegraphics [width=11cm]{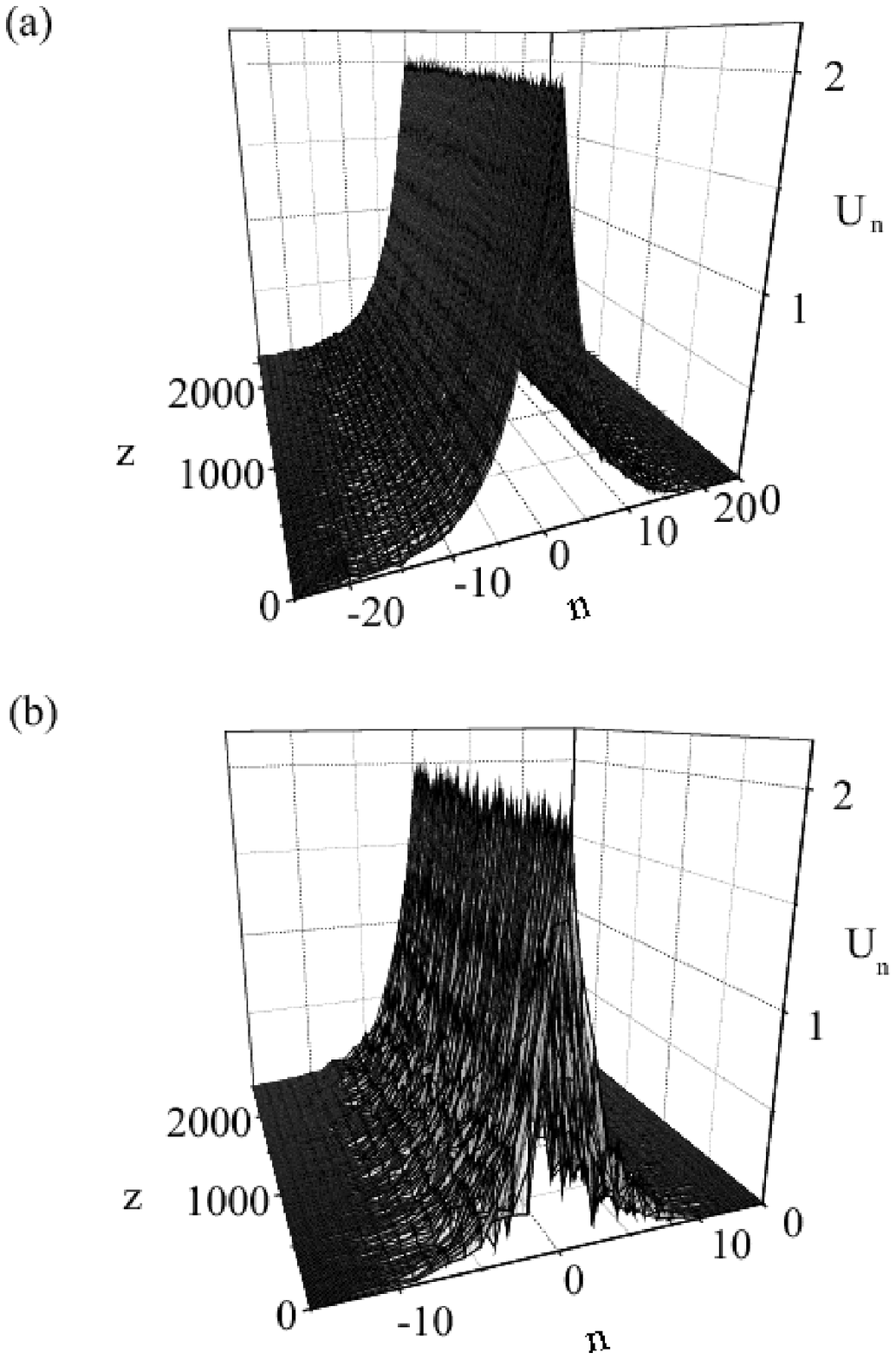}
\caption{The evolution of perturbed unstaggered solitons: (a) $\protect%
\alpha =0.5,\,K=-4$; (b) $\protect\alpha =1,\,K=-4$. The linear stability
analysis predicts the stability for (a) and instability for (b). }
\label{f4}
\end{figure}

Direct simulations confirm the actual stability of the UnST solitons
predicted by the linear-stability analysis, see Fig. \ref{f4}(a). As
concerns unstable solitons, they spontaneously evolve into confined
irregularly oscillating breathers, as shown in Fig. \ref{f4}(b). The
internal frequency of the breather is determined by the imaginary part of
the complex instability eigenvalues which govern the transformation of
stationary solitons into the breathers.

As shown in Fig. \ref{f1}(b), we have also found numerical solutions for the
twisted soliton modes in the infinite lattice, with $U_{-n}=-U_{+n}$, which
were introduced above by Eq. (\ref{inf-tru}). They coexist with the UnST and
ST solitons in the infinite lattice, but are always unstable, according to
the linear stability analysis and direct simulations (unlike the stability
of the similar solitons in the truncated lattice, see below). The
simulations (not shown here in detail) demonstrate that the instability
breaks the antisymmetry of those modes, as a consequence of a small
oscillating field amplitude appearing at $n=0$.

\subsection{The variational approximation}

\begin{figure}[h]
\center\includegraphics [width=7cm]{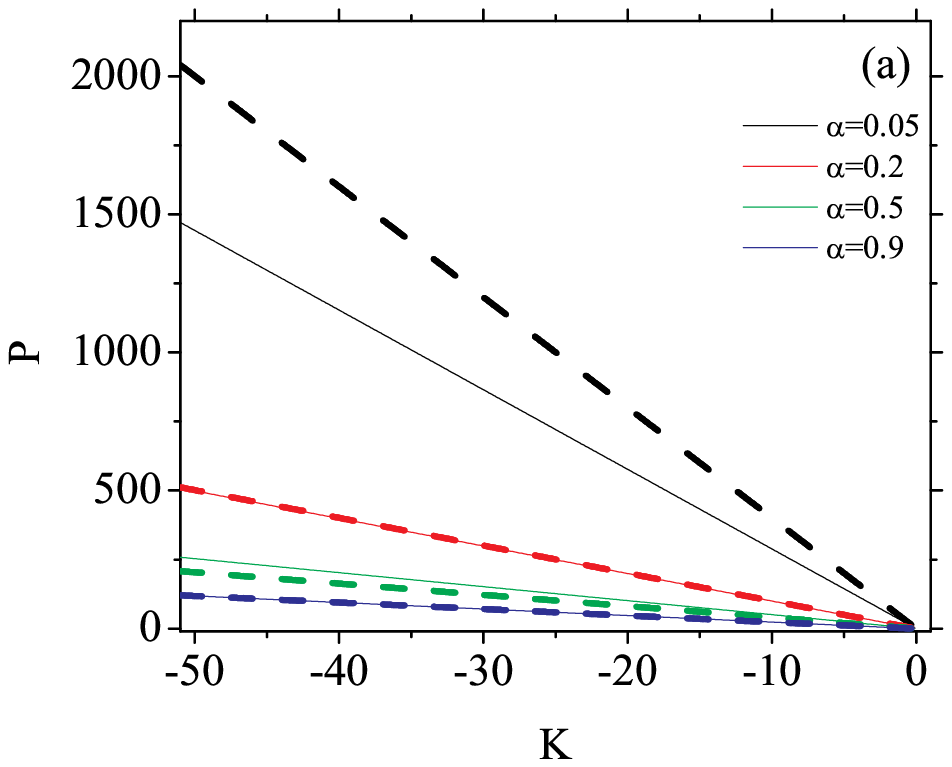}
\includegraphics
[width=7cm]{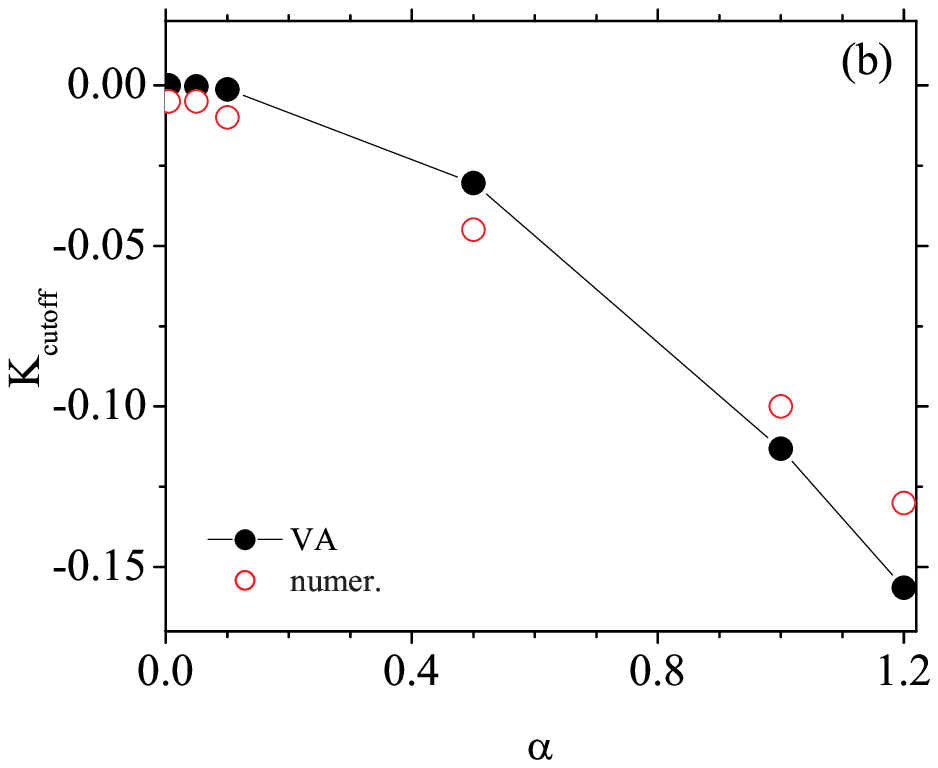} \caption{(Color online) (a) The $P$ vs.
$K$ dependencies for numerically found UnST solitons (solid
lines), and their counterparts predicted by the VA (dashed lines),
for fixed values of $\protect\alpha $ indicated in the panel. (b)
The corresponding cutoff values of $K$ at which the soliton
branches originate. Lines with filled circles depict the VA
prediction given by Eq. (\protect\ref{cutoff}), while hollow
circles represent the numerical results.} \label{poredjenje}
\end{figure}

The comparison with the continuous models elaborated in Refs. \cite%
{Barcelona} and \cite{Zeng}, as well as previous works dealing with discrete
solitons in homogeneous nonlinear lattices \cite%
{discr-VA,coupler,nashi-ponovo,RBlit,Shanghai}, suggests that it may be
relevant to develop a VA, based on the simple ansatz for the onsite-centered
solution of the UnST type, which emulates the exact non-soliton solution (%
\ref{exact}):%
\begin{equation}
U_{n}=A\exp \left( -\frac{\alpha }{2}|n|\right) ,  \label{ansatz}
\end{equation}%
where $A$ is treated as a variational parameter. The substitution of this
ansatz into Lagrangian (\ref{L}) easily leads to the corresponding effective
Lagrangian:%
\begin{equation}
L_{\mathrm{eff}}=\frac{A^{2}}{2\sinh \left( \alpha /2\right) }\left[ \cosh
\left( \alpha /2\right) \left( 1+K+\frac{1}{2}A^{2}\right) -1\right] .
\label{Leff}
\end{equation}%
The variational equation, following from this Lagrangian, $\partial L_{%
\mathrm{eff}}/\partial \left( A^{2}\right) =0$, predicts the amplitude of
the soliton,%
\begin{equation}
A_{\mathrm{VA}}^{2}=-K-\left[ 1-\mathrm{sech}\left( \alpha /2\right) \right]
,  \label{A}
\end{equation}%
cf. Eq. (\ref{Aexact}). Note that Eq. (\ref{A}) predicts a cutoff value of
the propagation constant,%
\begin{equation}
K_{\mathrm{cutoff}}^{\mathrm{(VA)}}=-\left[ 1-\mathrm{sech}\left( \alpha
/2\right) \right] <0,  \label{cutoff}
\end{equation}%
which is negative, unlike the positive cutoff produced by the exact
non-soliton solution (\ref{exact}). Finally, the substitution of this result
into expression for total power of ansatz (\ref{ansatz}),%
\begin{equation}
P_{\mathrm{ansatz}}=A^{2}\coth \left( \alpha /2\right) ,  \label{Pans}
\end{equation}%
yields the prediction of the VA for the total power as a function of the
propagation constant:%
\begin{equation}
P_{\mathrm{VA}}=-K\coth \left( \alpha /2\right) -\tanh \left( \alpha
/4\right) .  \label{PVA}
\end{equation}%
In Fig. \ref{poredjenje}(a), we compare, at different fixed values of $%
\alpha $, the power-versus-$K$ curves of the numerically generated UnST
soliton families and their VA-predicted counterparts. It is seen that the
agreement, quite naturally, improves, with the increase of the
nonlinearity-modulation parameter $\alpha $, for tighter self-trapped modes.
Additionally, Fig. \ref{poredjenje}(b) compares the numerically found and
VA-predicted [see Eq. (\ref{cutoff})] cutoff values of the propagation
constant at which the UnST-soliton branches originate.

\section{Surface solitons in truncated lattices}

\subsection{Exact solutions and the variational approximation}

As mentioned above, a remarkable peculiarity of the truncated-lattice model,
based on Eqs. (\ref{m1})-(\ref{01}), is a possibility to find particular
solutions for surface solitons in an exact analytical form, which was not
possible for the infinite lattice [recall the exact solution given by Eqs. (%
\ref{exact}) and (\ref{Aexact}) is not appropriate for solitons].

\subsubsection{The exact solution for unstaggered surface solitons}

First, an exact solution for a UnST soliton pinned to the edge of the
lattice is looked for as
\begin{eqnarray}
U_{n} &=&A\exp \left( -\alpha n/2\right) ,~\mathrm{at}~~n\geq 1,~  \notag \\
U_{n} &=&U_{0}~\mathrm{at}~~n=0,  \label{Um}
\end{eqnarray}%
where $U_{0}$ may be different from $A$, cf. the exact solution for the
model with the uniform onsite SDF nonlinearity and exponentially decaying
inter-site couplings, given above by Eqs. (\ref{tilde})-(\ref{C0}). The
substitution of ansatz (\ref{Um}) into Eqs. (\ref{m1}) and (\ref{11}) yields
the following relations:%
\begin{equation}
K=\cosh \left( \alpha /2\right) -1-A^{2}.  \label{b}
\end{equation}%
\begin{equation}
U_{0}=A/C_{0}.  \label{U0}
\end{equation}%
Then, the substitution of the ansatz and relations (\ref{b}), (\ref{U0})
into Eq. (\ref{01}) yields the final result:
\begin{equation}
A^{2}=\frac{C_{0}^{2}}{2}\frac{e^{\alpha /2}+\left( 1-C_{0}^{2}\right)
e^{-\alpha /2}}{C_{0}^{2}-\sigma _{0}}.  \label{A^2}
\end{equation}%
As usual, the exact solution is the exceptional one, which can be found with
for the single value of the propagation constant, given by Eq. (\ref{b}),
like exceptional exact soliton solutions found in similar continuous models
\cite{Barcelona,Zeng}. Nevertheless, this exact solution is more generic
than the above-mentioned one represented by Eqs. (\ref{tilde})-(\ref{U0tilde}%
), because its existence does not require a special selection of the
coupling constant $C_{0}$, unlike condition (\ref{C0})

The exact solution is meaningful if Eq. (\ref{A^2}) yields $A^{2}>0$, i.e. $%
C_{0}^{2}$ belongs to either of the two intervals:%
\begin{eqnarray}
\sigma _{0} &<&C_{0}^{2}<1+e^{\alpha },~~\mathrm{if~}~\sigma
_{0}<1+e^{\alpha };  \label{exist1} \\
1+e^{\alpha } &<&C_{0}^{2}<\sigma _{0},~\mathrm{if~}~\sigma _{0}>1+e^{\alpha
}.  \label{exist2}
\end{eqnarray}%
Note that the nonlinearity increases monotonously from the edge into
the bulk of the lattice if condition $\sigma _{0}<e^{\alpha }$
holds, which excludes the existence interval (\ref{exist2}). In
addition, in the absence of the ``intersite-coupling anomaly" ,
i.e., for $C_{0}=1$, the remaining existence condition
(\ref{exist1}) implies that the
nonlinearity at the edge site, $n=0$, must be weaker than in the bulk: $%
\sigma _{0}<1$.

It is also relevant to mention that, in the same case of $C_{0}=1$,
amplitude (\ref{A^2}) of the exact solution never vanishes, in agreement
with the fact that the total power of surface solitons, created in uniform
truncated lattices, cannot be smaller than a finite threshold value \cite%
{molina}. On the other hand, admitting $C_{0}>1$ opens the way for vanishing
of threshold: indeed, expression (\ref{A^2}) vanishes at $C_{0}=\sqrt{%
1+e^{\alpha }}$.

\subsubsection{The exact solution for staggered surface solitons}

It is also possible to construct a particular exact solution to Eqs. (\ref%
{m1})-(\ref{01}) in the form of an ST soliton, although it turns out to be
unstable, as shown below. We demonstrate this solution here for the sake of
the completeness of the analysis.

The exact ST soliton is looked for as
\begin{eqnarray}
U_{n} &=&A(-1)^{n}\exp \left( -\alpha n/2\right) ,~\mathrm{at}~~n\geq 1,~
\notag \\
U_{n} &=&U_{0}~\mathrm{at}~~n=0,  \label{stagg}
\end{eqnarray}%
cf. Eq. (\ref{Um}). The substitution of this ansatz into the stationary
equations leads to the following results for the exact solution, which
replace Eqs. (\ref{b}) and (\ref{A^2}) obtained above for the UnST soliton:
\begin{equation}
K=-\left[ \cosh \left( \alpha /2\right) +1+A^{2}\right] ,  \label{b-st}
\end{equation}%
\begin{equation}
A^{2}=-\frac{C_{0}^{2}}{2}\frac{e^{\alpha /2}+\left( 1-C_{0}^{2}\right)
e^{-\alpha /2}}{C_{0}^{2}-\sigma _{0}},  \label{A^2-st}
\end{equation}%
while relation (\ref{U0}) remains the same as before. As it follows from Eq.
(\ref{A^2-st}), the exact solution is meaningful (giving $A^{2}>0$) exactly
in those regions where the above exact solution for the UnST soliton \textit{%
does not} exist, cf. Eqs. (\ref{exist1}) and (\ref{exist2}). Being
interested in the case of $\sigma _{0}<e^{\alpha }$, when the strength of
the onsite SDF grows monotonously from $n=0$ towards $n\rightarrow \infty $,
we finally conclude that the exact solution for the ST soliton exists at $%
C_{0}^{2}<\sigma _{0}$. In particular, in the case of $C_{0}=1$ (no
``coupling anomaly" at the surface), this solution exists under
condition $\sigma _{0}>1$ (in fact, in interval $1<\sigma
_{0}<e^{\alpha }$, according to what is said above), which is
precisely opposite\textit{\ } to\ the above-mentioned existence
condition, $\sigma _{0}<1$, for the exact unstaggered solution, in
the same case of $C_{0}=1$.

\subsubsection{The variational approximation for unstaggered surface solitons%
}

Because the existence of UnST solitons is the most essential feature of the
model with the spatially modulated onsite SDF nonlinearity, and the above
exact solution is available at the single value of the propagation constant,
given by Eq. (\ref{b}), it makes sense to apply the VA to the description of
generic solutions for the UnST solitons. Recently, the VA was applied to
discrete surface solitons in truncated lattices with the homogeneous onsite
nonlinearity \cite{RBlit,Shanghai}.

We here aim to develop the VA for the surface solitons in the basic
model with $C_{0}=\sigma _{0}=1$, without ``anomalies" at the edge
of the lattice, when the corresponding Lagrangian (\ref{lagrangian})
reduces
to%
\begin{equation}
L=\frac{1}{2}\sum_{n=0}^{\infty }\left[ \left( K+1\right)
U_{n}^{2}-U_{n}U_{n+1}+\frac{1}{2}e^{\alpha n}U_{n}^{4}\right] .
\label{simple}
\end{equation}%
Essentially the same ansatz (\ref{ansatz}), as used above for discrete
solitons in the infinite lattice,
\begin{equation}
U_{n}=A\exp \left( -\left( \alpha /2\right) n\right) ,~n\geq 0,  \label{surf}
\end{equation}%
yields the following result, upon the substitution into Eq. (\ref{simple}):%
\begin{equation}
L_{\mathrm{eff}}=\frac{A^{2}}{4\sinh \left( \alpha /2\right) }\left[ \exp
\left( \alpha /2\right) \left( 1+K+\frac{1}{2}A^{2}\right) -1\right] ,
\end{equation}%
cf. Eq. (\ref{Leff}). In the present case, the variational equation, $%
\partial L_{\mathrm{eff}}/\partial \left( A^{2}\right) =0$, yields%
\begin{equation}
A_{\mathrm{VA}}^{2}=-\left[ K+1-\exp \left( -\alpha /2\right) \right] ,
\label{AVA}
\end{equation}%
cf. Eq. (\ref{A}), with the respective cutoff at%
\begin{equation}
K_{\mathrm{cutoff}}^{\mathrm{(surf-VA)}}=-\left[ 1-\exp \left( -\alpha
/2\right) \right] <0,  \label{cutoff-surf}
\end{equation}%
cf. Eq. (\ref{cutoff}). Finally, the total power of ansatz (\ref{surf}), $P_{%
\mathrm{ansatz}}^{\mathrm{(surf)}}=A^{2}\left[ 1-\exp \left( -\alpha \right) %
\right] ^{-1}$ [cf. Eq. (\ref{Pans})], yields the following prediction for
the $P(K)$ curve for the family of UnST surface solitons:%
\begin{equation}
P_{\mathrm{VA}}^{\mathrm{(surf)}}=-K\left[ 1-\exp \left( -\alpha \right) %
\right] ^{-1}-\left[ 1+\exp \left( -\alpha /2\right) \right] ^{-1},
\label{PVA-surf}
\end{equation}%
cf. the result (\ref{PVA}) predicted by the VA for the UnST solitons in the
infinite lattice. Comparison of the analytical results for $P_{\mathrm{VA}}^{%
\mathrm{(surf)}}$ and $K_{\mathrm{cutoff}}^{\mathrm{(surf-VA)}}$, which are
given by Eqs. (\ref{PVA-surf}) and (\ref{cutoff-surf}), with the numerically
obtained counterparts is presented in Figs. \ref{figure1a}(c) and (d),
respectively.

\subsection{Numerical results}

In the truncated lattice with the homogenous SDF nonlinearity, $\alpha
=0,\,\sigma _{0}=\sigma =1$, and uniform coupling between the lattice sites,
$C_{0}=1$, the existence of two distinct families of discrete solitons of
the ST type, pinned at the interface, one stable and one unstable, is well
known (see, e.g., Ref. \cite{molina}). These solitons exist at the total
power exceeding a finite threshold value, which is necessary for the
lattice-pinning force to overcome the repulsion from the surface.

In the present model with the inhomogeneous nonlinearity, numerical results
demonstrate that a nearly completely stable ST branch continues to exist for
all values of $\alpha $, $C_{0}$, and $\sigma $, starting with a finite
threshold power. These solitons are strongly pinned to the edge of the
lattice, very weakly depending on the modulation steepness $\alpha $, and
are stable in their almost entire existence region. However, comparison of
the numerically generated ST solitons with the exact solution given by Eqs. (%
\ref{stagg})-(\ref{A^2-st}) demonstrates that the latter one and numerically
found stable ST surface solitons belong to \emph{different} solution
branches. Accordingly, the linear-stability analysis shows that the exact
solution is unstable (not shown here in detail).

\begin{figure}[h]
\center\includegraphics [width=13cm]{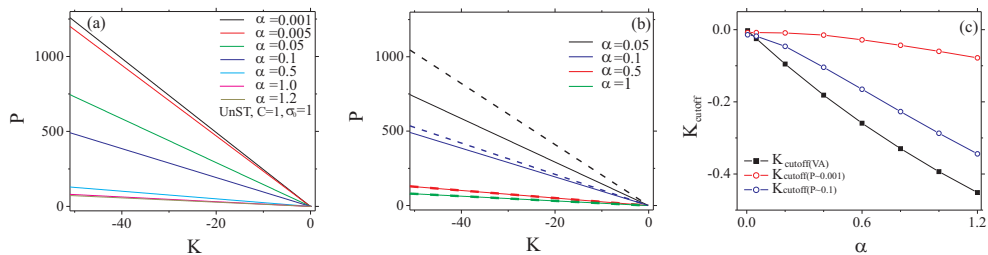}
\caption{(Color online) (a) The total power as a function of the
propagation
constant for unstaggered surface solitons. In plot (b), numerically found $%
P(K)$ curves (solid lines) and their VA-predicted counterparts given by Eq. (%
\protect\ref{PVA-surf}) (dashed lines) are compared for fixed values of $%
\protect\alpha $. Comparison of the prediction (\protect\ref{cutoff-surf})
for $K_{\mathrm{cutoff}}^{\mathrm{(surf-VA)}}$ with the corresponding
numerically found cutoff value of the propagation constant is shown in (c).
The red (upper) and blue (lower) curves corresponds to the cutoff defined by
setting the total power of the numerically found surface solitons to $%
P_{\min }=0.001$ and $P_{\min }=0.1$, respectively. Here, parameters are $%
C_{0}=1$ and $\protect\sigma _{0}=1$.}
\label{figure1a}
\end{figure}

\begin{figure}[h]
\center\includegraphics [width=13cm]{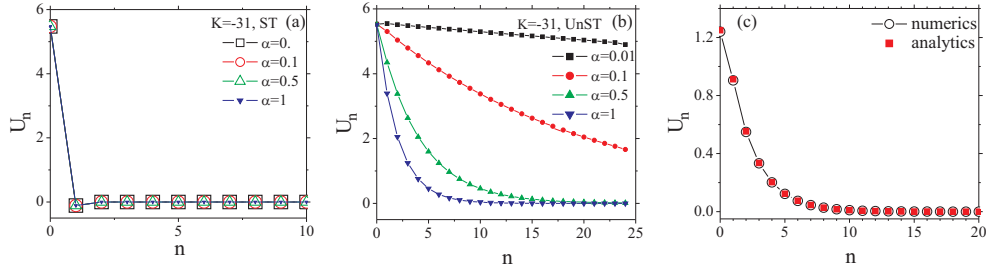}
\caption{(Color online) (a) and (b): Profiles of staggered (a) and
unstaggered surface solitons (b), coexisting at $C_{0}=1.2$ and $\protect%
\sigma _{0}=1$. Values of other parameters are indicated in the panels. (c)
Profiles of the exact analytical solution for the unstaggered surface
soliton and its numerically found counterpart (their coincidence verifies
that the analytical and numerical solutions belong to the same branch), for $%
k=1.13$, $C_{0}=1.2$, and $\protect\sigma _{0}=1$. }
\label{figuresur}
\end{figure}

The most essential numerical results concern the existence and stability of
UnST solitons centered at the lattice surface. We have performed this
analysis for various values of parameters $C_{0}$ and $\sigma _{0}$, and for
nonlinearity-modulation steepness taking values in the interval of $0<\alpha
<1.5$, where stable UnST surface solitons can be found (as well as in the
previous section, this limitation is imposed by the numerical scheme used,
rather than by the system itself). The $P(K)$ curves for this family, and
examples of solitons profiles are shown in Fig. \ref{figure1a}. It is worthy
to mention that the so found UnST surface solitons coexist with their
numerically generated counterparts of the ST type in a broad parameter
range, as shown in Fig. \ref{figuresur} for $\sigma _{0}=1$, $C_{0}=1.2$,
and different values of $\alpha $ (recall that the particular exact
analytical solutions for both types of the surface solitons do not coexist,
as shown above).

A noteworthy novel property of the UnST solitons is vanishing of the
threshold value of the total power necessary for their existence in the
basic model $C_{0}=\sigma _{0}=1$, see Fig. \ref{figure1a}(c) and (d). This
finding agrees with the variational results (\ref{AVA}) and (\ref%
{cutoff-surf}), which predict that the threshold is zero for the UnST
solitons. The absence of the threshold has obvious implications for the
physical realizations of the system, strongly facilitating the creation of
the solitons.

Unlike the ST surface solitons, whose shape is almost independent of the
nonlinearity-modulation rate $\alpha $, the shape of the UnST solitons is
quite sensitive to $\alpha $. As illustrated in Fig. \ref{figuresur}(c), in
all cases when the particular exact solution for the UnST soliton, given by
Eqs. (\ref{Um})-(\ref{A^2}), exists, it is identical to the numerical
solution found at the same parameters (on the contrary to the
above-mentioned case of the ST solitons, where the actually found stable
numerical solutions and the unstable analytical one belong to different
branches).

The numerical analysis demonstrates that the UnST surface solitons are
stable almost everywhere in their existence region. The respective
perturbation spectrum contains a small number of the complex eigenvalues
with a finite real part only in a narrow area close to the existence
threshold (not shown here in detail). Dynamical simulations of the perturbed
evolution confirm the stability of the UnST solitons. In particular, the
exact analytical soliton of the UnST type, which fits its numerical
counterpart in Fig. \ref{figuresur} (b), is stable too.

\section{Conclusion and extensions}

It is commonly known that the only localized modes in lattices
with the homogeneous SDF (self-defocusing) nonlinearity are
represented by ST (staggered) solitons. Here, we have demonstrated
that infinite and semi-infinite (truncated) lattices support
stable discrete UnST (unstaggered) solitons too, provided that the
strength of the onsite SDF nonlinearity grows rapidly enough from
the center towards the periphery. The physical implementation of
this setting may be provided by decreasing the detuning of the
underlying resonance (the two-photon one in optics, or the
Feshbach resonance in the BEC) in the same direction, from the
center to periphery. Families of the UnST solitons have been found
in the numerical form, and also reproduced by the VA (variational
approximation). In addition to that, particular exact solutions
were found for the UnST surface solitons in the semi-infinite
lattice. Stability regions for the UnST solitons have been found
by means of the calculation of eigenvalues for perturbation modes,
and verified by means of direct simulations. A noteworthy result
is vanishing of the threshold value of the total power (norm)
necessary for the existence of the surface UnST solitons in the
truncated lattice. In both the infinite and truncated ones, the
stable UnST solitons coexist with the (usual) stable ST-soliton
families. The settings considered here may be implemented for
matter waves in BEC trapped in deep optical lattices, and for
light waves in arrays of optical waveguides. In addition to that,
we have also introduced and briefly considered the lattice with
the uniform SDF nonlinearity and exponentially decaying inter-site
coupling. Physically, the latter feature can be readily provided
by gradual increase of the lattice spacing. Discrete UnST solitons
may exists in this system too, despite the SDF sign of the
nonlinearity.

This work can be naturally extended in other directions. In particular, it
may be interesting to study a similar system based on a set of two parallel
linearly coupled lattices, i.e., a discrete nonlinear coupler, which is
known in both infinite \cite{coupler} and semi-infinite \cite{Shanghai}
forms.

A challenging problem is to generalize the analysis for two-dimensional
(2D)\ lattices. In particular, a natural object may be a \textit{%
corner-shaped} lattice (cf. Ref. \cite{corner}), with the SDF onsite
nonlinearity modulated so that the stationary version of the 2D discrete
nonlinear Schr\"{o}dinger equation takes the following form [cf. Eq. (\ref%
{eq2})]:%
\begin{eqnarray}
-\left( K+2\right) U_{m,n} &=&-\frac{1}{2}\left(
U_{m+1,n}+U_{m-1,n}+U_{m,n+1}+U_{m,n-1}\right) +\sigma e^{\alpha \left(
m+n\right) }U_{m,n}^{3},~\mathrm{at}~~m,n\geq 1,  \notag \\
-\left( K+2\right) U_{m,0} &=&-\frac{1}{2}\left(
U_{m+1,0}+U_{m-1,0}+U_{m,1}\right) +\sigma _{0}e^{\alpha m}U_{m,0}^{3},~%
\mathrm{at}~~m\geq 1,n=0,  \notag \\
-\left( K+2\right) U_{0,n} &=&-\frac{1}{2}\left(
U_{0,n+1}+U_{0,n-1}+U_{1,n}\right) +\sigma _{0}e^{\alpha n}U_{0,n}^{3},~%
\mathrm{at}~~n\geq 1,m=0,  \notag \\
-\left( K+2\right) U_{0,0} &=&-\frac{1}{2}\left( U_{1,0}+U_{0,1}\right)
+\sigma _{00}U_{0,0}^{3},~\mathrm{at}~~m=n=0.  \label{corner}
\end{eqnarray}%
It is assumed that linear couplings are uniform throughout the 2D lattice,
but the nonlinearity coefficient may be modified near edges of the corner ($%
\sigma _{0}\neq \sigma )$, and additionally at the corner site ($\sigma \neq
\sigma _{00}\neq \sigma _{0}$). Then, it is easy to demonstrate, as a proof
of principle, that Eqs. (\ref{corner}) admit a particular exact solution for
2D UnST\textit{\ corner soliton}:%
\begin{equation}
U_{m,n}=A\exp \left( -\frac{\alpha }{2}\left( m+n\right) \right) ,
\end{equation}%
\begin{equation}
A^{2}=\left( \sigma -\sigma _{00}\right) ^{-1}\exp \left( \frac{\alpha }{2}%
\right) ,~K=2\cosh \left( \frac{\alpha }{2}\right) -2-\sigma A^{2},
\end{equation}%
under additional conditions%
\begin{equation}
\sigma +\sigma _{00}=2\sigma _{0},\sigma _{00}<\sigma .  \label{>}
\end{equation}

Moreover, an exact solution for a ST corner soliton can be found too:%
\begin{equation}
U_{m,n}=A(-1)^{n}\exp \left( -\frac{\alpha }{2}\left( m+n\right) \right) ,
\end{equation}%
\begin{equation}
A^{2}=\left( \sigma _{00}-\sigma \right) ^{-1}\exp \left( \frac{\alpha }{2}%
\right) ,~K=-2\cosh \left( \frac{\alpha }{2}\right) -2-\sigma A^{2}.
\end{equation}%
In the latter case, the additional condition is opposite to that given by
Eq. (\ref{>}) for the UnST mode:
\begin{equation}
\sigma +\sigma _{00}=2\sigma _{0},\sigma <\sigma _{00}.
\end{equation}%
The 2D system will be considered in detail elsewhere.

\acknowledgments G.G., A.M., and Lj.H. acknowledge support from the Ministry
of Education, Science and Technological Development of the Republic of
Serbia (Project III45010). B.A.M. appreciates a valuable discussion with Y.
V. Kartashov, and hospitality of the Vin\v{c}a Institute (Belgrade).

\end{document}